# Nonlinear dynamics of spinning fluid-conveying pipes with structural damping: stability analysis and post-instability behavior

Ali Fasihi[a,b], Grzegorz Kudra[a*], Maryam Ghandchi Tehrani[b], Jan Awrejcewicz[a]

[a] *Department of Automation, Biomechanics, and Mechatronics, Lodz University of Technology, 1/15 Stefanowski St., Lodz, 90-537, Poland.*

[b] *Dynamics and Vibration Group, Faculty of Science and Engineering, University of Groningen, Nijenborgh4, Groningen, 9747 AG, Netherlands.*

[*] *Corresponding author: grzegorz.kudra@p.lodz.pl*

## Abstract

Nonlinear dynamics of fluid conveying pipe, rotating with constant velocity about its longitudinal axis is analyzed. Considering boundary conditions and internal damping, the nonlinear equation of motion is derived, and it is discretized via the Galerkin method. Afterward, the stability of the system is investigated by characterizing the eigenvalues under the action of two control parameters: rotational speed and flow velocity. Then using direct numerical simulation, instability and stability regions are distinguished in a map as the control parameters vary. It is shown that due to the presence of internal damping in the system, both rotational speed and flow velocity determine the critical speeds. Post-instability behavior is characterized by non-zero equilibrium points, representing deflection in the rotating frame, which correspond to forward whirling motion in the inertial frame. Finally, Hencky Bar-chain Model was employed to verify the results.

## 1. Introduction

The study of fluid-conveying pipe dynamics has been one the most well-known within the field of the dynamics and stability of mechanical structures in the past decades, due to both its wide range of applications and its capability to show rich dynamical behavior. The substantial research dedicated to this field proves its importance. Païdoussis's comprehensive book [1], and his recent review article [2], provide an in-depth overview of the extensive work done on this topic. Two key phenomena—buckling and flutter instabilities—dominate the dynamics of these systems, driven by Coriolis and centrifugal forces generated by fluid flow. At the same time, rotordynamics is another paradigm in the dynamics of mechanical structures, where rotation similarly induces Coriolis and centrifugal forces. Foundational knowledge in rotordynamics can be found in the monograph by Ishida and Yamamoto [3], while Genta's book [4] provides a review of specialized subjects typically found only in journal articles. The intersection of these two fields—fluid-conveying pipes and rotordynamics—leads to the study of spinning pipes conveying fluid, a more intricate mechanical system subjected to both fluid flow and spinning motion. This area of research is recent, having emerged within the past few years, starting with research on drill string dynamics, where investigators gradually recognized that the interaction between the drill string and drilling fluid could be one of the important sources causing severe vibrations, leading to fatigue damage in the drill pipe [5]. This recognition prompted researchers to develop mathematical models that characterize drill string-mud interaction. For example, Zhang and Miska [6] conducted a theoretical study on the buckling and lateral vibrations of drill pipes in vertical wells interacting with incompressible fluids. Their findings revealed that there exists a critical flow rate at which the pipe becomes unstable due to either divergence (buckling) or flutter, and that this critical flow rate depends on factors such as system geometry, boundary conditions, axial forces, and the properties of both the pipe and fluid. Subsequently, Païdoussis and colleagues [6] extended this work by investigating the effects of both internal and external mud flow on drill strings, finding that the sensitivity of these effects was strongly influenced by the annular space between the drill string and the wellbore. Building on these insights, Eftekhari and Hosseini [7] explored the stability of cantilevered spinning pipes made from functionally graded materials, subjected to fluid and thermomechanical loading. More recently, in a series of studies, Liang et al. investigated the stability and dynamics of spinning pipes with different types of supports [8–11]. However, all of these analyses are based on linear models.

The equations of motion derived in the aforementioned studies are based on infinitesimal strain theory, leading to linear equations that are valid only for small amplitude motion. However, when oscillation amplitudes become large, the assumption of linearity breaks down, necessitating the inclusion of nonlinear terms—at least up to the third order—to account for geometrical nonlinearity. To study post-instability dynamical behavior, which involves large deflections, a nonlinear model is essential. For slender mechanical structures, the approximation of geometrical nonlinearity in the equations of motion depends on whether the structure can be considered extensible or inextensible [12]. For both cases, nonlinear models of fluid-conveying pipes have been developed, and their post-bifurcation behavior has been widely studied [13–15]. However, for spinning pipes, existing nonlinear models are limited to two studies [16,17], and none of them have explored the post-instability dynamics of the system.

In modeling conservative fluid-conveying pipes (such as pinned-pinned or clamped-clamped configurations), dissipative forces do not affect the stability region, leaving it independent of damping [18]. However, in rotating systems, damping can significantly alter this behavior. Specifically, when dealing with a damped rotor, it is crucial to distinguish between damping effects associated with the stationary parts of the machine, known as nonrotating damping, and those linked to the rotating parts, or rotating damping. The former typically provides a stabilizing influence that can be utilized to maintain stability across the machine's entire working range. The latter, however, may either stabilize or destabilize the system. As a result, the inclusion of structural damping plays a crucial role in the stability of spinning systems, contrasting sharply with cases where it is neglected [19].

The literature survey reveals that, to date, no studies have explored the post-instability dynamics of spinning pipes conveying fluid. This gap has motivated the present study, particularly to investigate the potential destabilizing effects of viscoelastic damping. The paper is structured as follows: in Section 2, the nonlinear equation of motion for the system is derived in its PDE form. Section 3 employs the Galerkin method to discretize the PDE into a set of ODEs. Section 4 discusses the results, and Section 5, to validate the developed nonlinear model, develops a Hencky model is introduced, representing the system as a series of rigid segments interconnected by universal joints.

## 2. Problem statement and equation of motion

Fig. 1 illustrates the system's schematic representation: a pipe with a length of $L$ and made of a material with a Young's modulus of $E$ which has a mass per unit length of $m$ and is supported by hinges at its ends. The pipe spins at a constant rotational speed of $\Omega$ about its axial direction and conveys fluid inside that has a mass per unit length of $M$ and a relative constant flow velocity ($U$) compared to the pipe.

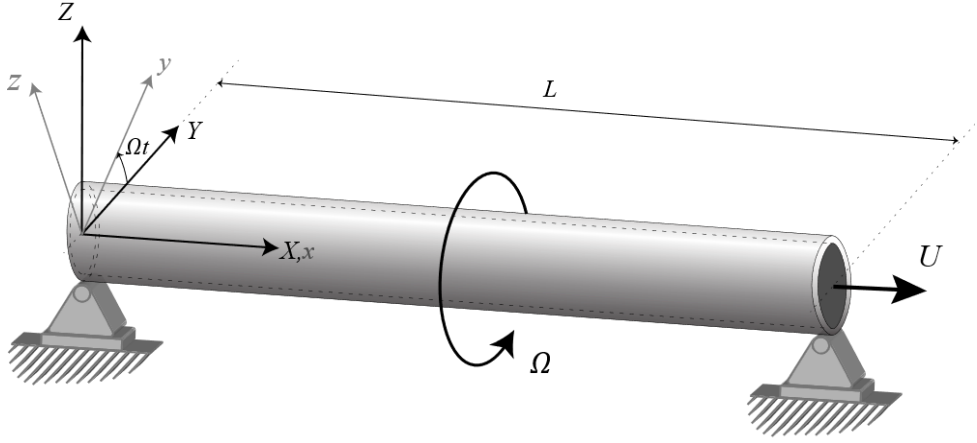

Fig. 1. Schematic representation of the system and its corresponding coordinates

Two coordinate systems are shown in Fig. 1; an inertial frame denoted by (X, Y, Z), and a rotating frame (x, y, z). Using the rotating frame is a common strategy in the literature to simplify the calculation of position vector. This is because it avoids the complications caused by the spinning motion ($\Omega$) when calculating the displacements. However, it's important to remember that the resulting equations and the motion and the corresponding frequencies they provide are for the rotating frame, not for the fixed one. Based on this strategy, the position vector of an arbitrary point of the pipe (*a*) at any instant is written in the rotating frame as

$$\mathbf{r}_p = (a_x + u)\,\mathbf{i} + v\,\mathbf{j} + w\,\mathbf{k}, \tag{1}$$

where $a_x$ is the x-component of the position of the point before deformation which defines which point of the pipe along its length we are referring to, and $u$, $v$, and $w$ are the displacements of the point in $x$, $y$, and $z$ direction, respectively, with $\mathbf{i}$, $\mathbf{j}$, and $\mathbf{k}$ being the unit vectors for each direction. Since the direction of these unit vectors change in time, they have time derivatives as

$$\dot{\mathbf{i}} = 0, \quad \dot{\mathbf{j}} = \Omega \mathbf{k}, \quad \dot{\mathbf{k}} = -\Omega \mathbf{j}. \tag{2}$$

Using Eqs. 1 and 2, the velocities of the pipe as well as fluid inside the pipe are given by

$$\begin{aligned}\dot{\mathbf{r}}_p &= (\dot{u})\,\mathbf{i} + (\dot{v} - \Omega w)\,\mathbf{j} + (\dot{w} + \Omega v)\,\mathbf{k},\\ \dot{\mathbf{r}}_f &= \bigl(\dot{u} + U(1+u')\bigr)\,\mathbf{i} + (\dot{v} - \Omega w + Uv')\,\mathbf{j} + (\dot{w} + \Omega v + Uw')\,\mathbf{k},\end{aligned} \tag{3a,b}$$

where the dots and primes denote $\frac{\partial}{\partial t}$ and $\frac{\partial}{\partial x}$, respectively. Having determined the velocities of the fluid and the pipe, the total kinetic energy of the system can now be calculated from Eq. (4)

$$T = T_p + T_f = \tfrac{1}{2}\, m \int_0^L (\dot{\mathbf{r}}_p{}^T \dot{\mathbf{r}}_p)\,dx + \tfrac{1}{2}\, M \int_0^L (\dot{\mathbf{r}}_f{}^T \dot{\mathbf{r}}_f)\,dx. \tag{4}$$

The general formula for potential energy is written as

$$V = \tfrac{1}{2} \iiint_{V_p} \sigma \varepsilon \, dV_p, \tag{5}$$

where $V_p$ shows the volume of the pipe and $\sigma$ and $\varepsilon$ are stress and strain. Due to the fixed nature of the pipe ends (simply supported), they cannot move in any direction. As a result, deflections cause stretching in the pipe. In such cases, the strain is written as [20]

$$\varepsilon = e + a_z k_y - a_y k_z, \tag{6}$$

where $e$ is the strain along the neutral axis of the pipe and given by

$$e = \sqrt{(1+u')^2 + v'^2 + w'^2} - 1, \tag{7}$$

and $a_z$ and $a_y$ are z and y-component of the position of the point *a*; and $k_y$ and $k_z$ are component of curvature. To account for internal damping within the system, the Kelvin–Voigt model is employed.

According to this model, the viscoelastic property of the pipe material manifests its effect in the relationship between strain and stress.

$$\sigma = E\varepsilon + \eta\dot{\varepsilon}, \tag{8}$$

where $\sigma$ is stress, and $\eta$ coefficient of viscosity. Eqs. 6 and 8 are substituted into Eq.5, thereby obtaining the potential energy of the system as following.

$$V = (E + \eta \frac{\partial}{\partial x}) \left[ \frac{1}{2} A \int_0^L (u'^2 + u'v'^2 + u'w'^2) dx + \frac{1}{2} I \int_0^L (v''^2 + w''^2) dx \right]. \tag{9}$$

With the kinetic energy from Eq.4 and this potential energy, Hamilton's principle can now be applied. Performing the required variational operations and considerable manipulation, the governing equations for the two lateral directions are obtained.

$$\eta I \dot{v}'''' + EI v'''' + (M+m)\ddot{v} + 2MU\dot{v}' + MU^2 v'' - 2(M+m)\Omega\dot{w} - (M+m)\Omega^2 v - 2M\Omega U w' - \frac{EA}{2L} v'' \int_0^L (v'^2 + w'^2) dx - \frac{\eta A}{L} v'' \int_0^L (v'\dot{v}' + w'\dot{w}') dx = 0,$$

$$\eta I \dot{w}'''' + EI w'''' + (M+m)\ddot{w} + 2MU\dot{w}' + MU^2 w'' + 2(M+m)\Omega\dot{v} - (M+m)\Omega^2 w + 2M\Omega U v' - \frac{EA}{2L} w'' \int_0^L (v'^2 + w'^2) dx - \frac{\eta A}{L} w'' \int_0^L (v'\dot{v}' + w'\dot{w}') dx = 0. \tag{10a,b}$$

It worth to mention that first terms of both equations are related to damping. The second and third terms address the system's stiffness and inertia. The fourth and fifth terms represent the Coriolis and centripetal forces induced by fluid flow, respectively. The sixth and seventh terms correspond to the Coriolis and centripetal forces due to spinning motion. The eighth term is another gyroscopic effect, resulting from both of fluid flow and spinning motion. Finally, the ninth and tenth terms represent the nonlinear stiffness and damping, respectively, which arise from the stretching of the pipe.

To make the analysis more general, the following dimensionless parameters are introduced.

$$x^* = \frac{x}{L}, \ v^* = \frac{v}{L}, \ w^* = \frac{v}{L}, \ t^* = \left(\frac{EI}{M+m}\right)^{1/2} \frac{t}{L^2}, \ \alpha = \left(\frac{I}{E(M+m)}\right)^{1/2} \frac{\eta}{L^2}, \ \beta = \frac{M}{M+m}, \ U^* = \left(\frac{M}{EI}\right)^2 LU, \ \Omega^* = \left(\frac{M+m}{EI}\right)^{1/2} L^2 \Omega, \ \mu = \frac{AL^2}{I}. \tag{11}$$

Eqs. (10a,b) are rewritten in dimensionless form.

$$\alpha\dot{v}'''' + v'''' + \ddot{v} + 2\beta^{1/2} U \dot{v}' + U^2 v'' - 2\Omega\dot{w} - \Omega^2 v - 2\beta^{1/2} \Omega U w' - \frac{1}{2}\mu v'' \int_0^1 (v'^2 + w'^2) dx - \alpha\mu v'' \int_0^1 (v'\dot{v}' + w'\dot{w}') dx = 0,$$

$$\alpha\dot{w}'''' + w'''' + \ddot{w} + 2\beta^{1/2} U \dot{w}' + U^2 w'' + 2\Omega\dot{v} - \Omega^2 w + 2\beta^{1/2} \Omega U v' - \frac{1}{2}\mu w'' \int_0^1 (v'^2 + w'^2) dx - \alpha\mu w'' \int_0^1 (v'\dot{v}' + w'\dot{w}') dx = 0. \tag{12a,b}$$

To make the equation easier to read, the overbars of the quantities have been omitted. By defining the complex number $r = v + iw$, these equations can also be expressed in a single equation as follows,

$$\alpha\dot{r}'''' + r'''' + \ddot{r} + 2\beta^{1/2} U \dot{r}' + U^2 r'' + i2\Omega\dot{r} - \Omega^2 r + i2\beta^{1/2} \Omega U r' - \frac{1}{2}\mu r'' \int_0^1 (r'\bar{r}') dx - \frac{1}{2}\alpha\mu r'' \int_0^1 (r'\dot{\bar{r}}' + \bar{r}'\dot{r}') dx = 0, \tag{13}$$

where $\bar{r}$ is the complex conjugate of $r$. This complex form is particularly useful when investigating the post-divergence behavior of the system because its magnitude shows total deflection of the system with respect to its origin.

## 3. Galerkin Method

The Galerkin discretization method is one the most common method for analysing continuous systems, it provides a reduced-order model (ROM) of continuous systems by transforming their partial differential equations into a set of ordinary differential equations. According to this method, the solution of Eqs. 10 can be expressed as

$$v(x,t) = \sum_{j=1}^n \varphi_j(x) q_{vj}(t),$$
$$w(x,t) = \sum_{j=1}^n \varphi_j(x) q_{wj}(t), \tag{14a,b}$$

where $\varphi_j(x)$ are appropriate comparison functions satisfying the boundary conditions, $q_j(t)$ are the generalized coordinates of the discretized system, and $n$ the Galerkin truncation number indicating the number of modes taken into account. For the simply supported pipe, comparison functions for the pipe system are chosen to be

$$\varphi_i(x) = \sqrt{2} \sin j\pi x, \quad j = 1,2,\ldots,n. \tag{15}$$

By applying Galerkin method, the following set of ODEs are obtained.

$$\mathbf{M\ddot{q} + G\dot{q} + C\dot{q} + Kq + Nq = 0}, \tag{16}$$

where **M**, **G**, **C**, **K**, and **N** are mass, gyroscopic, damping, stiffness and nonlinearity matrices, respectively.

$$\mathbf{q} = \begin{bmatrix} \mathbf{q_v} \\ \mathbf{q_w} \end{bmatrix}, \quad \mathbf{M} = \begin{bmatrix} \mathbf{A} & 0 \\ 0 & \mathbf{A} \end{bmatrix}, \quad \mathbf{G} = \begin{bmatrix} 2\beta^{1/2}U\mathbf{B} & -2\Omega\mathbf{A} \\ 2\Omega\mathbf{A} & 2\beta^{1/2}U\mathbf{B} \end{bmatrix}, \quad \mathbf{C} = \begin{bmatrix} \alpha\mathbf{D} & 0 \\ 0 & \alpha\mathbf{D} \end{bmatrix}, \quad \mathbf{K} =$$

$$\begin{bmatrix} \mathbf{D} + u^2\mathbf{C} - \Omega^2\mathbf{A} & -2\beta^{1/2}U\Omega\mathbf{B} \\ 2\beta^{1/2}U\Omega\mathbf{B} & \mathbf{D} + u^2\mathbf{C} - \Omega^2\mathbf{A} \end{bmatrix}, \quad \mathbf{N} = \begin{bmatrix} -\mu\mathbf{C}(\frac{1}{2}\mathbf{E} + \alpha\mathbf{F}) & 0 \\ 0 & -\mu\mathbf{C}(\frac{1}{2}\mathbf{E} + \alpha\mathbf{F}) \end{bmatrix}$$

$$\begin{aligned} A_{ij} &= \int_0^1 \varphi_j \varphi_i \, dx \\ B_{ij} &= \int_0^1 \varphi_j' \varphi_i \, dx \\ C_{ij} &= \int_0^1 \varphi_j'' \varphi_i \, dx \\ D_{ij} &= \int_0^1 \varphi_j'''' \varphi_i \, dx \\ E_i &= \int_0^1 \sum_{j=1}^n (\varphi_j')^2 (q_{vj} + q_{wj})^2 \, dx \\ F_i &= \int_0^1 \sum_{j=1}^n (\varphi_j')^2 (q_{vj}\dot{q}_{vj} + q_{wj}\dot{q}_{wj}) \, dx \end{aligned} \tag{17}$$

## 4. Results

The mathematical model developed above is used this section to explore the characteristics and properties of the system's dynamics. This is done using direct numerical simulation and frequency analysis. Direct numerical simulation involves direct time integration of discretized equations and calculating solutions for a given initial condition using the Runge-Kutta technique. Frequency analysis involves calculating the eigenvalues of the linearized system to determine the natural frequencies of its modes and assess the system's stability. In general, to find natural frequencies of the system, periodic solutions of the linearized version of the discretized form of EOMs, are sought to be

$$\mathbf{q} = \mathbf{p}e^{i\omega t}, \tag{18}$$

where $i$ is the imaginary unit, **p** is a column of unknown amplitudes, and $\omega$ is the natural frequency. Introducing it into the leads to the following generalized eigenvalue problem.

$$(-\omega^2 \mathbf{M} + i\omega\,\mathbf{G} + i\omega\mathbf{C} + \mathbf{K})\mathbf{p} = 0, \tag{19}$$

The determinant of the coefficient matrix should be zero to obtain the nontrivial solution of the above equation.

$$\det(-\omega^2 \mathbf{M} + i\omega\,\mathbf{G} + i\omega\mathbf{C} + \mathbf{K}) = 0. \tag{20}$$

In rotordynamics literature, the governing equation of motions and consequently their solutions typically are obtained in the inertial frame. However, the equation of motions derived above is described in the rotating frame and not inertial frame. Using the following transformation the equation of motions is obtained in the inertial frame.

$$r = \rho e^{-i\Omega t}, \tag{21}$$

where, $r$ is the complex number as defined in Eq. 13, and $\rho$ is its counterpart in the inertial frame. Calculating its first- and second-time derivatives as

$$\dot{r} = (\dot{\rho} - i\Omega\rho)e^{-i\Omega t},$$
$$\ddot{r} = (\ddot{\rho} - 2i\Omega\dot{\rho} + \Omega^2\rho)e^{-i\Omega t}, \tag{22a,b}$$

and substituting them into Eq. 13, EOM in the inertial frame is obtained as follows.

$$\alpha\dot{\rho}'''' - i\alpha\Omega\rho'''' + \rho'''' + \ddot{\rho} + 2\beta^{1/2} U \dot{\rho}' + U^2 \rho'' - \frac{1}{2}\mu\rho'' \int_0^1 (\rho'\bar{\rho}') \, dx - \frac{1}{2}\alpha\mu\rho'' \int_0^1 (\rho'\dot{\bar{\rho}}' + \dot{\rho}'\bar{\rho}' - i2\Omega\rho'\bar{\rho}') \, dx = 0 . \tag{23}$$

It is observed that two terms (the second and the final term) are related to the spinning motion, and both include damping coefficients. These terms are associated with the destabilizing effect of rotating damping. The rotational speed determines their magnitudes and thus their effect on the results, ultimately determining whether the system is stable or unstable. This will be discussed later in Section 4.2. Another observation is that the transformation causes all the terms related to spinning motion in the rotating frame to disappear in the inertial frame. Thus, if internal damping is not considered, there is no rotational speed ($\Omega$) in the inertial frame EOMs. This means that spinning motion has no effect on the characteristics of the undamped system, and only the FSI determines them. Knowing this, it is natural to ask about the effect of spinning motion terms found in the rotating frame equation: How do they alter the solutions, and how can these changes be interpreted? To address these questions, the problem is simplified by removing flow velocity related terms from both rotating frame and inertial frame EOMs, and their results are compared.

### 4.1 Rotating and inertial frame results

Starting with the inertial frame EOMs, discretizing them using the Galerkin method, the equations are obtained in matrix form (similar to Eq. 16). Solving it for eigenvalues according to the procedure explained in (Eq. 18-20), natural frequencies corresponding to the first mode of motion are obtained as follows.

$$_{(I)}\omega_{1^{st}\, mode} = \pm \pi^2 \tag{24}$$

The prefix subscript $I$ in this expression indicates that these frequencies are calculated in the inertial frame. In the vibration of beam-like structures, including fluid-conveying pipes, which experience flexural vibration, negative roots of the characteristic equation are not considered as frequencies and are omitted, as they are mathematically obtained values with no physical meaning. However, this is not the case here, since the frequencies obtained above are not flexural but angular frequencies. This will be explained further below. The matrix form of the equations of motion for one-mode Galerkin truncation is

$$\begin{bmatrix} 1 & 0 \\ 0 & 1 \end{bmatrix} \begin{bmatrix} \ddot{q}_{1V} \\ \ddot{q}_{1W} \end{bmatrix} + \begin{bmatrix} \pi^4 & 0 \\ 0 & \pi^4 \end{bmatrix} \begin{bmatrix} q_{1V} \\ q_{1W} \end{bmatrix} = 0, \tag{25}$$

where $q_{1V}$ and $q_{1W}$ represent displacements in $Y$ and $Z$ direction of the inertial frame. Since there is no coupling term between these two equations, each of them is treated as a 1 DOF system.

$$\ddot{q}_{1V} + \pi^4 q_{1V} = 0,$$
$$\ddot{q}_{1W} + \pi^4 q_{1W} = 0. \tag{26a,b}$$

and therefore, have closed-form solutions as

$$q_{1V} = A\sin(\pi^2 t + \alpha),$$
$$q_{1W} = B\sin(\pi^2 t + \beta). \quad (27a,b)$$

Each expression describes a free undamped vibration. The vibrations in the *Y* and *Z*-directions are independent of each other but have the same frequency, forming an orbit in the *YZ*-plane. In rotordynamics, this orbit is considered as a superposition of circular motions (i.e. forward and backward whirling motions). To express Eq. 27 as the sum of the two circular motions, $\alpha$ is set to 0 and $\beta$ is set to $\pi/2$, so it is rewritten as

$$q_{1V} = A\sin\pi^2 t,$$
$$q_{1W} = B\cos\pi^2 t. \quad (28a,b)$$

These equations represent the elliptical motion showed on the left side in Fig. 2. By rearranging them, one can write

$$q_{1V} = \frac{A+B}{2}\sin\pi^2 t + \frac{A-B}{2}\sin(-\pi^2 t),$$
$$q_{1W} = \frac{A+B}{2}\cos\pi^2 t + \frac{A-B}{2}\cos(-\pi^2 t). \quad (29a,b)$$

From this expression, it is observed that the elliptical motion is composed of two circular motions with the same magnitude of $\pi^2$, but moving in opposite directions (shown in the rest of Fig. 2). The motion in a counterclockwise direction has an angular velocity of $\pi^2$ and a radius of $(A+B)/2$, and the clockwise motion has an angular velocity of $-\pi^2$ and a radius of $|A-B|/2$. Since the counterclockwise motion matches the direction of the pipe's rotation ($\Omega$), it is referred to as forward whirling motion, whereas the clockwise motion, which is opposite to the pipe's rotation direction, is called backward whirling motion.

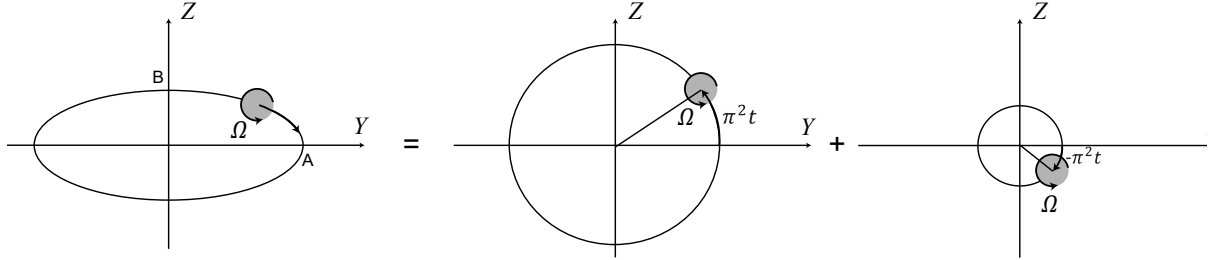

*Fig. 2.; Decomposition of a whirling motion.*

So, in general, the positive roots of the characteristic equation are natural angular of frequencies of forward whirling motions (or shortly forward whirling frequencies), and the negative roots of the characteristic equation are backward whirling frequencies. Having analysed the inertial frame EOMs, the rotating frame equations of motion are now considered. Due to the presence of spinning motion terms in the rotating frame equation, the natural angular frequencies corresponding to the first mode of motion are obtained as follows.

$$_{(R)}\omega_{1st\ mode} = \pm\pi^2 - \Omega \quad (30)$$

The prefix subscript *R* denotes that these frequencies are in the rotating frame. When comparing these natural frequencies with those obtained in the inertial frame, it is observed that the forward whirling in the rotating frame has a lower frequency than the forward whirling in the inertial frame by the value of $\Omega$, as the rotating frame rotates in the same direction as the forward whirling motion. Conversely, for the backward whirling motion, which is opposite to the direction of rotation of the rotating frame, $\Omega$ is added to the frequency. The variation of natural frequencies with respect to the rotational speed (aka Campbell diagram) is plotted in **Fig. 3** for both the inertial frame EOMs and the rotating frame EOMs.

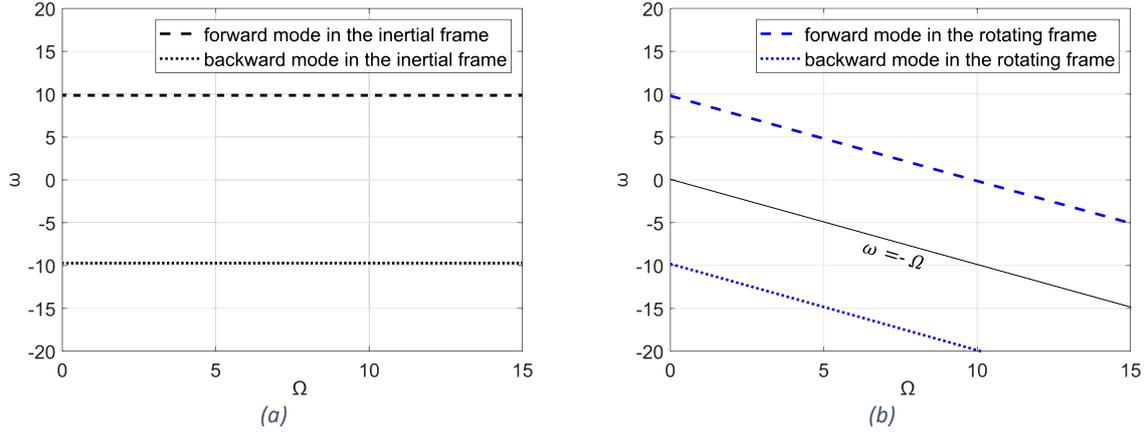

*Fig. 3. The natural frequency of the system versus rotational speed, Ω, in (a) inertial frame and in (b) the rotating frame; α = 0, U = 0.*

Fig. 3 (a) shows that increasing rational speed has no effect on the system's frequencies, while in the rotating frame, the frequencies of the system decrease as the rotational speed increases (as shown in **Fig. 3** (b)). It is noticed that at $\Omega = \pi^2$, the forward whirling frequency becomes zero, and for higher values of $\Omega$, It takes on a negative value. This occurs when the time required for the rotating frame to complete one rotation is less than the time needed for the system to complete a forward whirl. In other words, the rotating frame rotates faster than the forward whirling motion. Consequently, forward whirling motion, like backward whirling motion, appears in the clockwise direction in the rotating frame. In conclusion, for an undamped system, the rotational speed does not affect the system's dynamics, but it causes a difference in the representation of whirling motions in the rotating and inertial frames.

### 4.2 The effect of damping ($\alpha \neq 0$)

In what was discussed up to now there were no dissipation in the system. Frequency analysis shows that if damping is added, the natural frequencies of the system have a complex form. The real parts of them denote the natural frequencies themself, and the imaginary parts are proportional to damping. From Eq. 18, it is deduced that the imaginary part determines the decay or growth rate of the amplitude, where a positive value means the motion decays exponentially over time, and a negative value means the motion grows exponentially. Fig. 5 illustrates the changes in both the real and imaginary parts of each frequency separately with respect to rotational speed. Fig. 4 (*a*) shows the real parts, and due to the EOMs being in the rotating frame, similar to what is shown in **Fig. 3** (b), the frequencies decrease with increasing rotational speed. Fig. 4 (*b*) presents the imaginary parts, indicating that forward whirling motions of each mode change sign, transitioning from positive to negative at specific rotational speeds. These speeds are critical speeds as the amplitude of the motions grows beyond these values. The critical speeds for the first and second modes are $\Omega = \pi^2$ and $\Omega = 4\pi^2$, respectively.

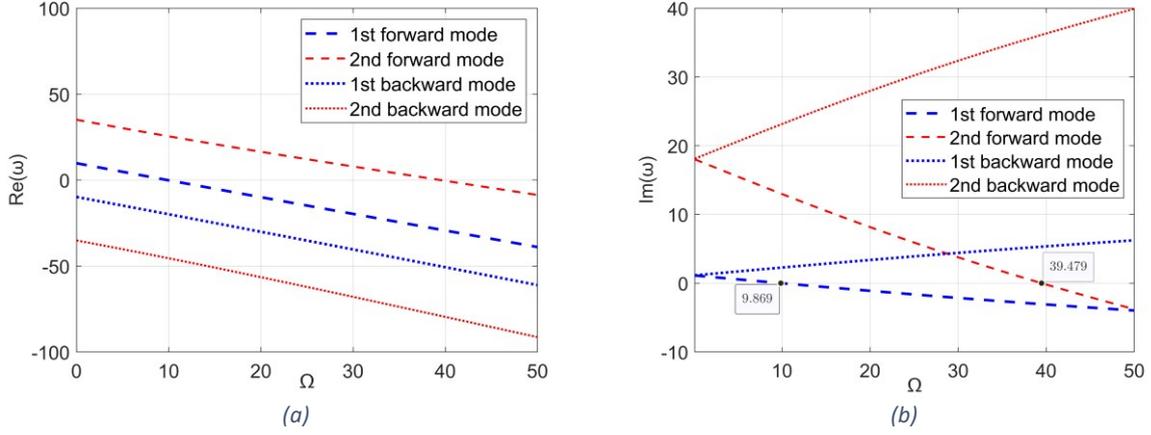

*Fig. 4. (a) the real and (b) imaginary components of the natural frequencies of the first two modes of system versus rotational speed, Ω, for the first two modes; α =0.02317, $U = 0$.*

Numerical simulation enables us to observe and quantify the growth and decay of amplitude for sub and supercritical speeds. Fig. 5 shows the time history of the midpoint of the system's length for two rotational speeds: one below and one above the critical speed of the first mode. In Fig. 5 (a), when the rotational speed is below the critical speed, the amplitude of motion decreases for any initial condition, leading to a steady-state solution of zero. Conversely, Fig. 5 (b) shows that when the rotational speed exceeds the critical value, a small perturbation causes the amplitude to grow, and it does not return to the zero-equilibrium position. However, due to the stiffness-hardening characteristic of the nonlinear term in the EOMs, the amplitude does not grow indefinitely but instead stabilizes at a non-zero equilibrium position, as shown for both the y and z directions.

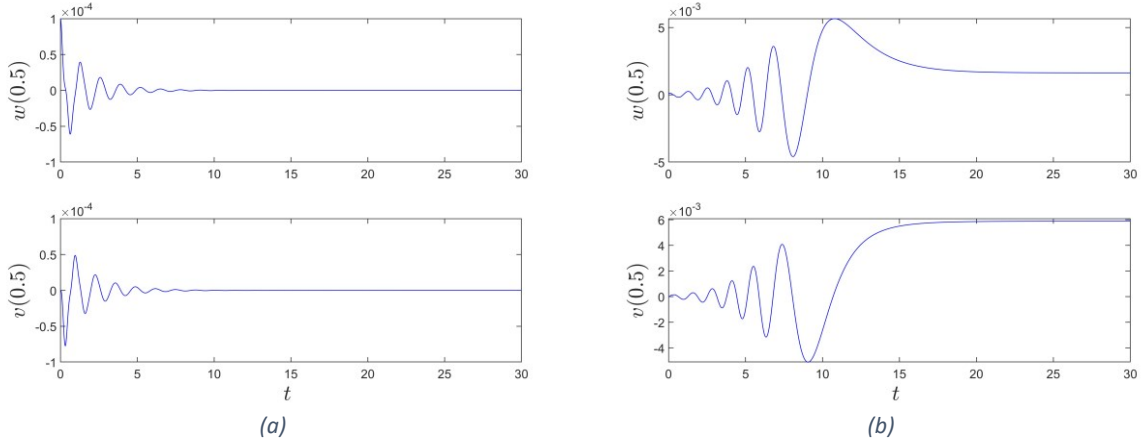

*Fig. 5. The time history of the midpoint of the system (a) below ($\Omega = 5$) and (b) above ($\Omega = 15$) the critical rotational speed; α = .02317, $U = 0$.*

These values are different between the two directions and may vary depending on the initial conditions. However, the distance of the non-zero equilibrium position from the zero equilibrium remains constant, regardless of the initial condition. This distance is defined as

$$|\mathrm{r}| = \sqrt{v^2 + w^2}, \tag{31}$$

which is the absolute value of r (introduced in Eq. 13) and represents the total deflection of the system. Fig. 6 shows the total deflection of the midpoint of the system with respect to rotational speed.

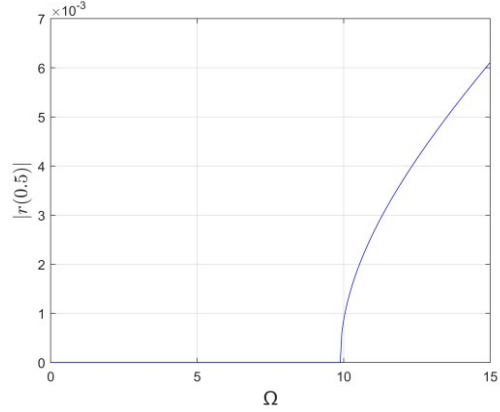

Fig. 6. Total deflection of the midpoint of the system with respect to rotational speed; α = 0.02317, $U = 0$.

As the figure shows, for rotational speeds below the critical value, total deflection is zero, and for rotational speeds higher than the critical value, the total deflection increases by increasing the rotational speed. It should be noted that the nonzero equilibrium point is the solution found for the rotating frame EOM. The corresponding solution in the inertial frame is not a nonzero equilibrium point but a forward whirling motion with the same magnitude of deflection. Video 1 in the supplementary data available online demonstrates this difference.

### 4.3 System with flow ($U \neq 0$)

With the effects of spinning motion and damping analyzed, we now turn to the full equations of motion in this section, incorporating the effects of flow velocity, to provide a more comprehensive analysis. for a given value of the rotational speed, we can perform frequency analysis and observe the changes in frequencies with respect to flow velocity. Two different values of rotational speed (i.e. $\Omega$ =5 and 8) were selected here, and the results are shown in Fig. 7 and Fig. 8. Both figures show that by increasing flow velocity, frequencies of all modes decrease parabolically, which is a well-known expected behavior of fluid-conveying pipes. Fig. 7 shows that at $U \approx 2.7$, the imaginary part of the forward whirling motion of the first mode changes sign from positive to negative, indicating that this mode becomes unstable at this speed. Similarly, the second mode becomes unstable at $U \approx 6.2$.

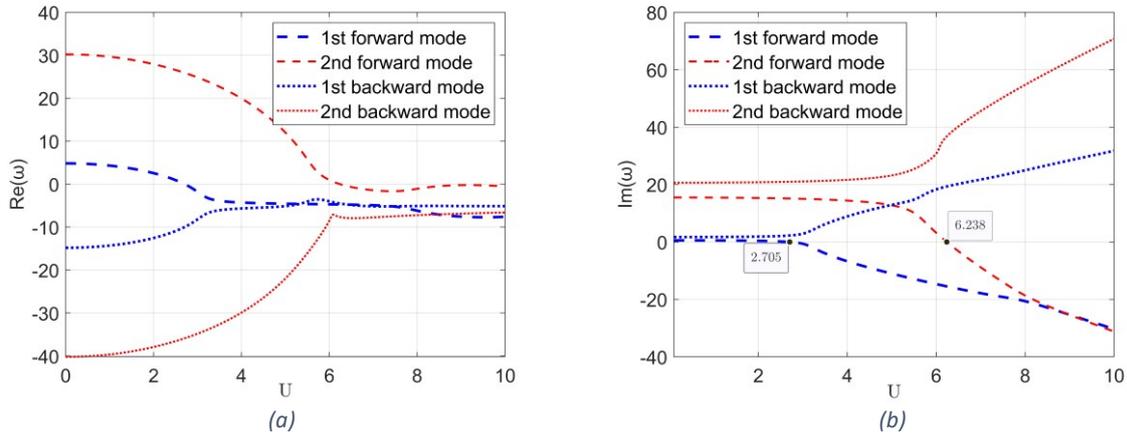

Fig. 7. (a) the real and (b) imaginary components of the natural frequency versus flow velocity, U, for the first two modes; α = 0.02317, $\Omega = 5$.

Fig. 8 is topologically similar to the previous figure, with the only difference being the critical speeds for the first and second modes, which occur at $U \approx 1.8$ and U $U \approx 6.1$, respectively. This indicates that as the rotational speed increases, the system loses stability at lower flow velocities.

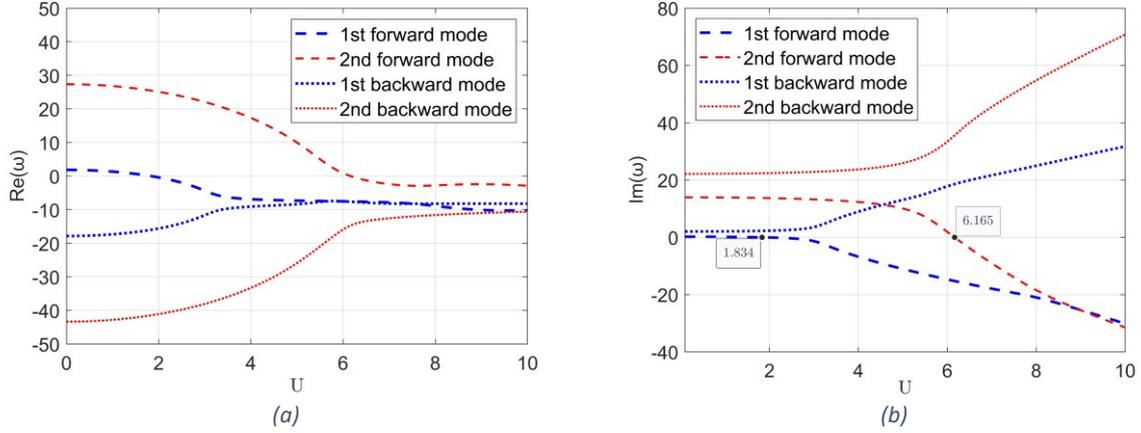

*Fig. 8. (a) the real and (b) imaginary components of the natural frequency versus flow velocity, U, for the first two modes; α = 0.02317, $\Omega = 8$.*

The numerical solution of the nonlinear equations also confirms this. Fig. 9 shows the total deflection of the midpoint of the system with respect to flow velocity: for $\Omega = 5$, the total deflection becomes non-zero for flow velocities above $U \approx 2.7$, while for $\Omega = 8$, the total deflection is non-zero for flow velocities above $U \approx 1.8$. The second mode's instability, which was predicted by the frequency analysis happens where the solution is already unstable and therefore does not materialize [21].

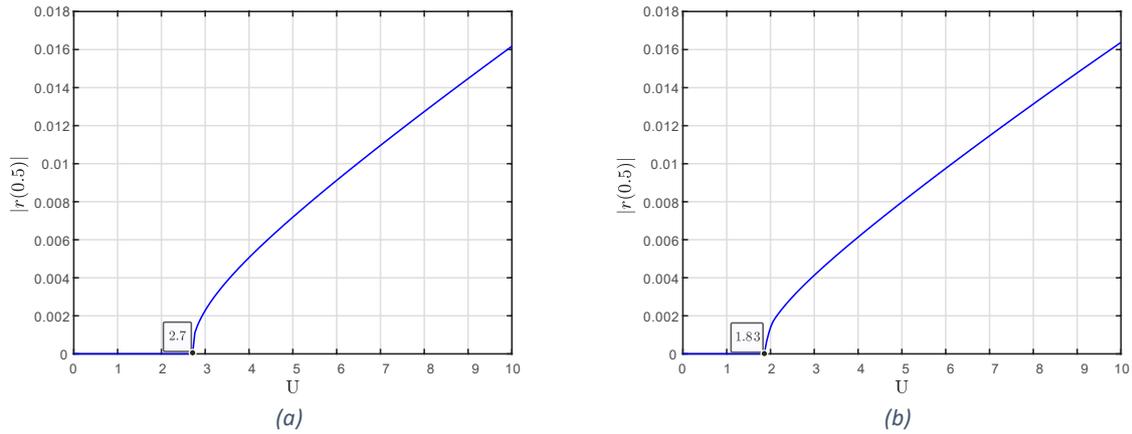

*Fig. 9. Total deflection of the midpoint of the system with respect to flow velocity; (a) $\Omega = 5$ and (b) $\Omega = 8$.*

To illustrate how the total deflection varies with both flow velocity and rotational speed, Fig. 10 presents a comprehensive view of these changes. When there is no spinning motion, the system degrades to a fluid-conveying pipe. The dynamics of this system are well-known and studied [1]. For this system, $U = \pi$ is the critical speed: below this value, any perturbation results in the system returning to its zero-equilibrium position, and above the critical speed, the stable solution is a non-zero point, corresponding to the buckling of the pipe. When rotational speed is introduced, the onset of divergence from the zero-equilibrium position occurs at progressively lower flow velocities as the rotational speed increases. In other words, the higher the rotational speed, the lower the flow velocity at which divergence begins. At $\Omega = \pi^2$, the system diverges even at zero flow velocity.

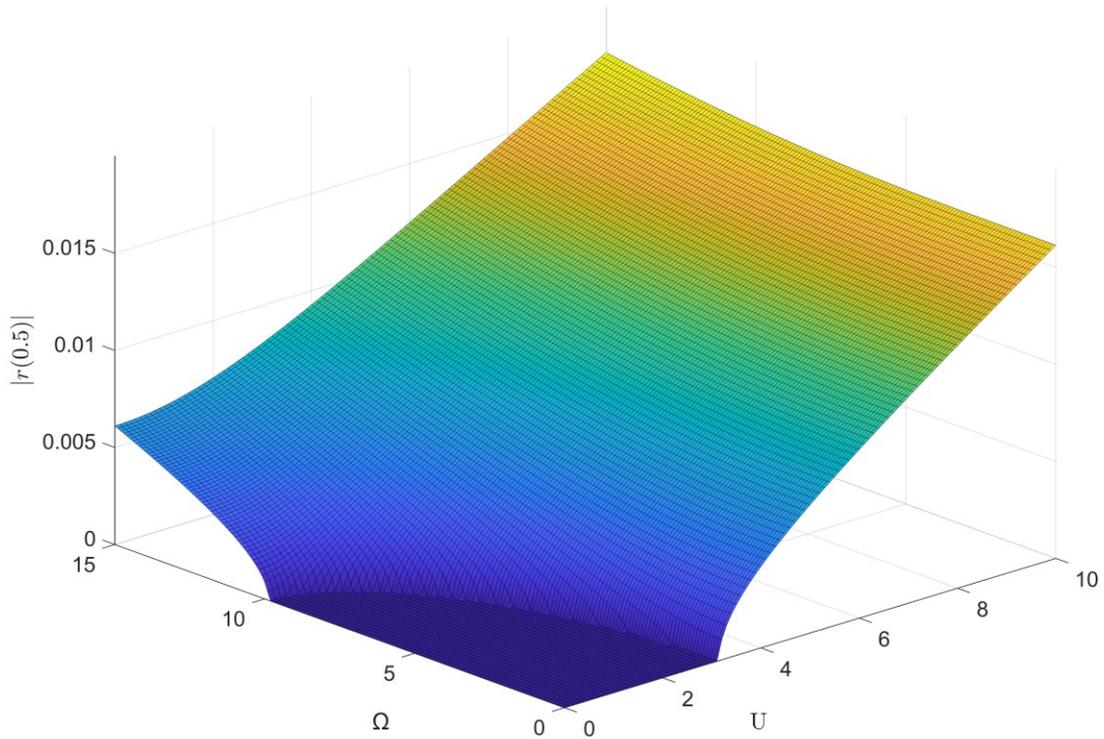

*Fig. 10. Total deflection of the midpoint of the system with respect to rotational speed and flow velocity.*

The left side of the figure where the flow velocity is zero resembles Fig. 6. Another important observation is that the non-zero equilibrium points grow larger with increasing *U* compared to increasing $\Omega$, indicating that flow velocity has a greater influence on the system's deflection than rotational speed. Finally, the region where there is no divergence and the zero-equilibrium position is stable is bounded by the critical flow velocity and critical rotational speed, forming an ellipse with a semi-major axis of $\pi$ and a semi-minor axis of $\pi^2$. This region is shown in Fig. 11.

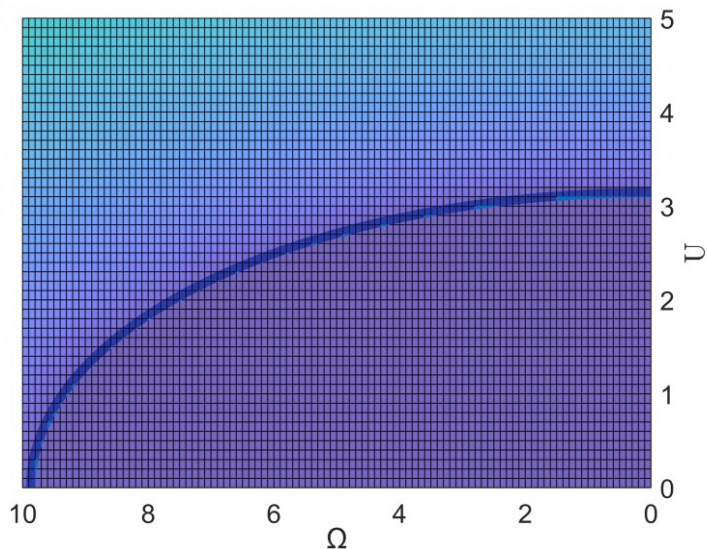

*Fig. 11. The ellipse that bounds the stable region of the system.*

## 5. Hencky Method

In this section, the Hencky Bar-chain Model, one of the oldest methods for studying continuous structures, is used to verify the model introduced in the previous section. In this approach, based on the continuous model of structures, a new segmented physical structural model is developed. This approach transforms the continuous system into a multi-rigid-body system. Having this considered, in a particular and original approach used in this work a new Hencky model is developed for the spinning pipe.

### 5.1 Developing the Hencky Model

First, the undeformed continuous pipe is divided into $n$ elements of equal length (Fig. 12 ($a$)). Next, in the center of each element there is placed a universal joint with specially selected angular springs and dampers intended to imitate the bending stiffness and internal damping of each pipe element. These elements, are infinitely thin and rigid hollow cylinders, rigidly connected to the neighbouring elements. Then, the first element's left end is connected via the same universal joint to the horizontally mounted shaft driving the pipe. The right end of the last element is supported elastically in three directions, with stiffness coefficients selected in such a way as to imitate the boundary conditions of the continuous pipe at its right end and at the same time simulate the stretching effect of continuous pipe in the Hencky model. And finally, a fluid flows at a constant relative velocity inside each element.

As a result of the discretization process described above, the system is transformed into a spatial configuration of $n + 1$ rigid links connected by universal joints. This forms an open kinematic chain with $2(n + 1)$ degrees of freedom. All the links in the system have the same length, except for the first and last links, which are each half the length of the remaining links. The moving links are numbered sequentially as $i = 0, 1, 2, \ldots, n$. The center of the universal joint, which coincides with the beginning of the $i$-th link, is denoted as $O_i$. Consequently, the end of the last link is marked as $O_{n+1}$. It is assumed that the joint sizes are negligibly small, and the links are treated as uniform bars, so the center of mass $C_i$ of each link lies at its geometric center. As with the continuous pipe, an inertial frame ($X, Y, Z$) is introduced, with its origin at point $O_0$. Additionally, body-attached frames ($x_i, y_i, z_i$) are introduced for each link $i = 0, 1, 2, \ldots, n$, with their origins at points $O_i$ (Fig. 12 ($b$)). A rotating frame ($x, y, z$), attached to the driving link and with its origin at point $O_0$, is also introduced. Each frame is equipped with an appropriate set of unit vectors ($\mathbf{u}_{x_i}, \mathbf{u}_{y_i}, \mathbf{u}_{z_i}$).

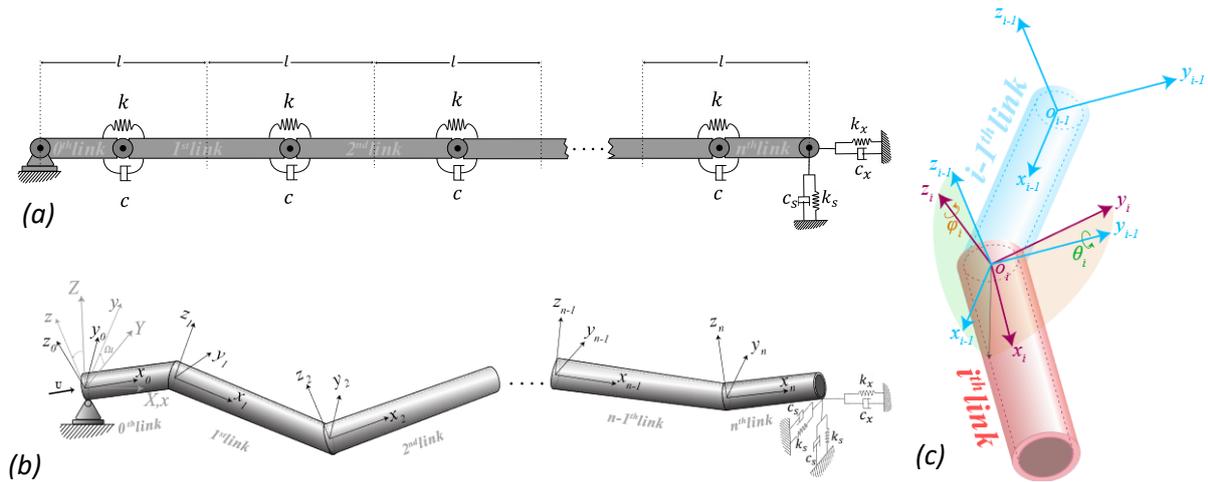

Fig. 12. (a) Schematic representation of the Hencky model of the system; (b) the corresponding coordinates of its links; and (c) relation between the coordinates of two consecutive links and their sequence of rotations.

The following rotation matrix defines the orientation of the rotating frame $x, y, z$ with respect to the inertial frame ($X, Y, Z$).

$$\mathbf{A}_0 = \begin{bmatrix} {}_{(I)}\mathbf{u}_x & {}_{(I)}\mathbf{u}_y & {}_{(I)}\mathbf{u}_z \end{bmatrix} = \begin{bmatrix} 1 & 0 & 0 \\ 0 & \cos \Omega t & -\sin \Omega t \\ 0 & \sin \Omega t & \cos \Omega t \end{bmatrix}, \tag{32}$$

where ${}_{(I)}\mathbf{u}_x$, ${}_{(I)}\mathbf{u}_y$ and ${}_{(I)}\mathbf{u}_z$ are the column vectors of the corresponding unit vectors of rotating frame, described in the inertial reference frame. It is clear, that rotating matrix $\mathbf{A}_0$ defines single rotation around the axis $X$ by an angle $\Omega t$. The orientation of each subsequent local reference frame is defined by the following rotation matrix.

$$\mathbf{B}_i = \begin{bmatrix} {}_{(i-1)}\mathbf{u}_{x_i} & {}_{(i-1)}\mathbf{u}_{y_i} & {}_{(i-1)}\mathbf{u}_{z_i} \end{bmatrix} = \mathbf{A}_{i1}\,\mathbf{A}_{i2}, \tag{33}$$

where

$$\mathbf{A}_{i1} = \begin{bmatrix} \cos \theta_i & 0 & \sin \theta_i \\ 0 & 1 & 0 \\ -\sin \theta_i & 0 & \cos \theta_i \end{bmatrix}, \quad \mathbf{A}_{i2} = \begin{bmatrix} \cos \varphi_i & -\sin \varphi_i & 0 \\ \sin \varphi_i & \cos \varphi_i & 0 \\ 0 & 0 & 1 \end{bmatrix}. \tag{34}$$

Here, ${}_{(i-1)}\mathbf{u}_{x_i}$, ${}_{(i-1)}\mathbf{u}_{y_i}$ and ${}_{(i-1)}\mathbf{u}_{z_i}$ are the column vectors of the corresponding unit vectors of $(x_i, y_i, z_i)$ reference frame, expressed in the reference frame of the previous link $(x_{i-1}, y_{i-1}, z_{i-1})$. For the case of $i = 0$, by ${}_{(-1)}\mathbf{u}_{x_0}$, ${}_{(-1)}\mathbf{u}_{y_0}$ and ${}_{(-1)}\mathbf{u}_{z_0}$ we understood the coordinates of the corresponding unit vectors in the rotating reference frame $(x, y, z)$. In other words $(x, y, z)$ is equivalent to $(x_{-1}, y_{-1}, z_{-1})$. Based on expressions 33 and 34, it is visible that the position of each subsequent link is obtained by two sequential rotations: a) by the $\theta_i$ angle around the axis $y_i$, b) by the $\varphi_i$ angle around the axis $z_i$ (Fig. 12 (c)).

Using the introduced angles describing the position of the system, the following vector of generalized coordinates is introduced as

$$\begin{aligned}\mathbf{q} &= [q_1 \quad q_2 \quad q_3 \quad q_4 \quad \ldots \quad q_{2(n+1)-1} \quad q_{2(n+1)}]^{\mathrm{T}} = \\ &\quad [\theta_0 \quad \varphi_0 \quad \theta_1 \quad \varphi_1 \quad \ldots \quad \theta_n \quad \varphi_n]^{\mathrm{T}}.\end{aligned} \tag{35}$$

The lumped model is created directly as a dimensionless model, corresponding to the dimensionless continuous model of pipe (Eqs. 12). Since according to this the pipe has a length equal to 1, the lengths of the individual links are

$$\begin{aligned} l_0 &= l_n = \tfrac{1}{2} l, \\ l_i &= l, \quad \text{for} \quad i = 1, 2, \ldots, n-1 \end{aligned} \tag{36}$$

where $l$ is the length of each internal link as well as the length of each element into which the pipe was divided during discretization.

$$l = \frac{1}{n}. \tag{37}$$

Since the total mass of the dimensionless pipe together with fluid is equal to 1, the nondimensional masses of the individual links $m_i$ and fluid $M_i$ inside them are

$$m_i = (1 - \beta^2) l_i, \quad M_i = \beta^2 l_i, \quad \text{for} \quad i = 0, 1, 2, \ldots, n. \tag{38}$$

Dimensionless stiffness coefficient of all angular springs in the individual universal joints is assumed as $k = (EI)^*/l$, where $(EI)^*$ is nondimensional flexural stiffness of pipe. Comparing dimensional (10) and dimensionless equations (12) of motion, one can easily see that $(EI)^* = 1$. By analogy, the angular damper coefficient mounted at each rotational joint is assumed to be $c = (\eta I)^*/l$, where $(\eta I)^* = \alpha$. Therefore the coefficients of angular springs and dampers read

$$k = \frac{1}{l}, \quad c = \frac{\alpha}{l}. \tag{39}$$

The stiffness and damping coefficients ($k_s$ and $c_s$) of transverse spring and dampers imitating the support of the right end of the pipe should be selected by simulation experiment, so that the right end of the pipe moves in the transverse direction to a negligible extend. However, they should not be too large due to the stability of the numerical simulation. Coefficients $k_x$ and $c_x$ of the spring and damper, at the right end of the pipe in the longitudinal direction, are introduced in order to simulate the stretching effect of continuous pipe in the Hencky model, because the latter is inextensible along the axes of its links. Therefore they should imitate the elasto-damping property of the pipe along its axis. Neglecting here deviation of the pipe from the horizontal direction, one can write $k_x = (EA)^*/L^*$ and $c_x = (\eta A)^*/L^*$, where $L^* = 1$ is dimensionless length of pipe, while $(EA)^*$ and $(\eta A)^*$ are dimensionless tensile stiffness and damping of the pipe, respectively. Comparing the corresponding terms in dimensional (10) and dimensionless equations (12), one can find that $(EA)^* = \mu$ and $(\eta A)^* = \alpha\mu$. Therefore the corresponding dimensionless coefficients read

$$k_x = \mu, \quad c_x = \alpha\mu. \tag{40}$$

Let us denote fluid velocity in the nondimensional Hencky model as $\tilde{U}$. This nondimensional velocity results directly from the relationships between dimensional and dimensionless time and length (11). Taking these relationships into account, in order to maintain the equivalence between the introduced dimensionless models, it turns out that

$$\tilde{U} = \frac{U}{\beta}, \tag{41}$$

where $U$ is parameter representing fluid velocity in nondimensional continuous model (Eqs. 12)

The position vectors $\mathbf{r}_{O_i} = \mathbf{r}_{O_i/O_0}$ of the joints $O_i$ are calculated sequentially according to the following relationship

$$\mathbf{r}_{O_{i+1}} = \mathbf{r}_{O_i} + \mathbf{r}_{\frac{O_{i+1}}{O_i}}, \quad \text{for } i = 0, 1, 2, \ldots, n. \tag{42}$$

In rotating reference frame $(x, y, z)$ it takes the form

$$_{(-1)}\mathbf{r}_{O_{i+1}} = {}_{(-1)}\mathbf{r}_{O_i} + \mathbf{B}_0 \mathbf{B}_1 \ldots \mathbf{B}_i {}_{(i)}\mathbf{r}_{\frac{O_{i+1}}{O_i}}, \quad \text{for } i = 0, 1, 2, \ldots, n, \tag{43}$$

where

$$_{(-1)}\mathbf{r}_{O_0} = [0 \quad 0 \quad 0]^T, \quad {}_{(i)}\mathbf{r}_{O_{i+1}/O_i} = [l_i \quad 0 \quad 0]^T. \tag{44}$$

In order to obtain the absolute position of a point $O_i$ in the inertial reference frame $(X, Y, Z)$, we use the relationship

$$_{(I)}\mathbf{r}_{O_i} = \mathbf{A}_{0\,(-1)}\mathbf{r}_{O_i}, \quad \text{for } i = 0, 1, 2, \ldots, n. \tag{45}$$

To develop the governing equations of motion for the system, Lagrange equations are utilized. To obtain the kinetic energy first, the linear and angular velocities of each link are calculated. For calculating absolute velocity $\mathbf{v}_{C_i}$ of mass center $C_i$ of each link, the following relations are used sequentially for each link $i = 0, 1, 2, \ldots, n$:

$$\mathbf{v}_{C_i} = \mathbf{v}_{O_i} + \boldsymbol{\omega}_i \times \mathbf{r}_{C_i/O_i}, \quad \mathbf{v}_{O_{i+1}} = \mathbf{v}_{O_i} + \boldsymbol{\omega}_i \times \mathbf{r}_{O_{i+1}/O_i}, \tag{46}$$

where $\mathbf{v}_{O_i}$ is the absolute velocity of point $O_i$, $\boldsymbol{\omega}_i$ is the absolute angular velocity of the $i$-th link while $\mathbf{r}_{C_i/O_i}$ is vector connecting the point $O_i$ with mass center $C_i$. The formulas (15) are utilized, and the components of the corresponding velocities are calculated sequentially in the local reference frames of the corresponding links:

$$_{(i)}\boldsymbol{\omega}_i = \mathbf{B}_i^T {}_{(i-1)}\boldsymbol{\omega}_{i-1} + \mathbf{A}_{i2}^T \begin{Bmatrix} 0 \\ \dot{\theta}_i \\ 0 \end{Bmatrix} + \begin{Bmatrix} 0 \\ 0 \\ \dot{\varphi}_i \end{Bmatrix}, \quad \text{for } i = 0, 1, 2, \ldots, n$$

$$_{(i)}\mathbf{v}_{O_i} = \mathbf{B}_i^T {}_{(i-1)}\mathbf{v}_{O_{i-1}}, \quad \text{for } i = 1, 2, 3, \ldots, n \tag{47}$$

where

$$_{(-1)}\boldsymbol{\omega}_{-1} = [\Omega \quad 0 \quad 0]^T, \quad _{(0)}\mathbf{v}_{O_0} = [0 \quad 0 \quad 0]^T, \quad _{(i)}\mathbf{r}_{C_i/O_i} = \begin{bmatrix} \frac{l_i}{2} & 0 & 0 \end{bmatrix}^T. \tag{48}$$

Therefore, the total kinetic energy of the system can be written as follow:

$$T = \frac{1}{2} \sum_{i=0}^{n} [m_i {}_{(i)}\mathbf{v}_{C_i}^T {}_{(i)}\mathbf{v}_{C_i} + {}_{(i)}\boldsymbol{\omega}_i^T {}_{(i)}\mathbf{I}_p {}_{(i)}\boldsymbol{\omega}_i + M_i {}_{(i)}\mathbf{v}_{f_i}^T {}_{(i)}\mathbf{v}_{f_i} + {}_{(i)}\boldsymbol{\omega}_i^T {}_{(i)}\mathbf{I}_f {}_{(i)}\boldsymbol{\omega}_i] \tag{49}$$

where ${}_{(i)}\mathbf{I}_p$ and ${}_{(i)}\mathbf{I}_f$ are inertia tensors of $i$-th link (pipe) and fluid inside it expressed the in the local reference frame $(x_i, y_i, z_i)$:

$$_{(i)}\mathbf{I}_p = \begin{bmatrix} 0 & 0 & 0 \\ 0 & \frac{m_i l_i^2}{12} & 0 \\ 0 & 0 & \frac{m_i l_i^2}{12} \end{bmatrix}, \quad _{(i)}\mathbf{I}_f = \begin{bmatrix} 0 & 0 & 0 \\ 0 & \frac{M_i l_i^2}{12} & 0 \\ 0 & 0 & \frac{M_i l_i^2}{12} \end{bmatrix}, \tag{50}$$

and where ${}_{(i)}\mathbf{v}_{f_i}$ is fluid velocity at the point $C_i$ in local reference frame:

$$_{(i)}\mathbf{v}_{f_i} = {}_{(i)}\mathbf{v}_{C_i} + \begin{bmatrix} \frac{U}{\beta} & 0 & 0 \end{bmatrix}^T. \tag{51}$$

Next, potential energy is calculated:

$$V = \frac{1}{2} k \sum_{i=1}^{n} \mathbf{q}^T \mathbf{q} + \frac{1}{2} {}_{(I)}\mathbf{r}_L^T \operatorname{diag}({}_{(I)}\mathbf{k}_L) {}_{(I)}\mathbf{r}_L, \tag{52}$$

where ${}_{(I)}\mathbf{k}_L$ is the matrix representing stiffness of the right end support of the pipe in the inertial reference frame

$$_{(I)}\mathbf{k}_L = [\mu \quad k_s \quad k_s]^T, \tag{53}$$

and where $\mathbf{r}_L = \mathbf{r}_{O_{n+1}} - \mathbf{r}_{O_{n+1}}\big|_{\mathbf{q}=0}$ is vector representing position of the right end of the pipe with respect to its initial position (for $\mathbf{q} = \mathbf{0}$):

$$_{(I)}\mathbf{r}_L = {}_{(I)}\mathbf{r}_{O_{n+1}} - [1 \quad 0 \quad 0]^T. \tag{54}$$

In order to take into account the dissipative forces, the following Rayleigh dissipative function is constructed

$$R = \frac{1}{2} c \sum_{i=1}^{n} \dot{\mathbf{q}}^T \dot{\mathbf{q}} + \frac{1}{2} {}_{(I)}\dot{\mathbf{r}}_L^T \operatorname{diag}({}_{(I)}\mathbf{c}_L) {}_{(I)}\dot{\mathbf{r}}_L, \tag{55}$$

where ${}_{(I)}\mathbf{c}_L$ is the matrix representing damping of the right end support of the pipe in the inertial reference frame

$$_{(I)}\mathbf{c}_L = [\alpha\mu \quad c_s \quad c_s]^T. \tag{56}$$

Using the introduced relations one can express governing differential equations of motion of the system based on Lagrange's equations:

$$\frac{d}{dt}\left(\frac{\partial T}{\partial \dot{q}_i}\right) - \frac{\partial T}{\partial q_i} + \frac{\partial V}{\partial q_i} + \frac{\partial R}{\partial \dot{q}_i} = F_i \quad i = 1,2,\ldots,n, \tag{57}$$

where $F_i$ is an additional generalized force accounting for the system's openness. An open system refers to one that experiences mass and momentum flow in and out at its boundaries. According to [22], it takes the form:

$$F_i = -M^* \widetilde{U}(\dot{\mathbf{r}}_L + \widetilde{U}\mathbf{u}_{x_n}) \cdot \frac{\partial \mathbf{r}_L}{\partial q_i}, \quad i = 1,2,\ldots,n, \tag{58}$$

where $M^* = M_i/l_i = M_r^2$ is the mass of the fluid per unit length and $\widetilde{U} = U/M_r$ is the fluid velocity in the introduced nondimensional Hencky model. Taking these relations into account and expressing the formula (27) in the inertia reference frame, one gets

$$F_i = -\beta U \left({}_{(I)}\dot{\mathbf{r}}_L + \frac{U}{M_r} {}_{(I)}\mathbf{u}_{x_n}\right)^T \frac{\partial {}_{(I)}\mathbf{r}_L}{\partial q_i}, \quad i = 1,2,\ldots,n, \tag{59}$$

where

$$\overset{...}{_{(I)}}\boldsymbol{u}_{x_n} = \mathbf{A}_0\mathbf{B}_0\mathbf{B}_1 \ldots \mathbf{B}_n[1 \quad 0 \quad 0]^{\mathrm{T}}. \tag{60}$$

Wolfram Mathematica environment has been used to implement all the formulas described above, and the resulting differential equations of motion (Eq. 57) can be solved for any chosen initial condition. The code written in Wolfram Mathematica 13.1 is available online in the supplementary data.

### 5.2 Comparison with the Galerkin model

With the differential equations of motion now obtained, we can move on to comparing the results of the Hencky and Galerkin models. For this purpose, the Hencky model is configured with five links (n=5). The rationale behind selecting this particular number of links will be addressed in the next subsection (section 5.3).

The first comparison is made under conditions of zero flow velocity and rotational speed, focusing solely on the nonlinearity from the stretching effect. As shown in Fig. 13 (a), for small amplitude perturbations, the basic Hencky model (without the longitudinal damper and spring elements at the right end) aligns well with the Galerkin model. However, Fig. 13 (b) demonstrates that for large amplitude perturbations, lateral displacement causes pipe extension, leading to significant differences between the basic Hencky model and the Galerkin model. Adding the spring element improves the frequency response of the Hencky model, bringing it closer to that of the Galerkin model, but the decaying behavior remains poorly captured. When the damper is added, both models come into very close agreement.

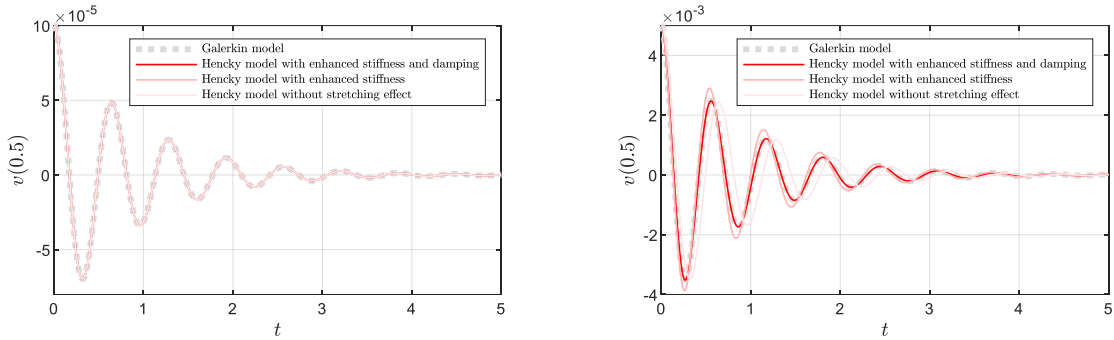

Fig. 13. Time history of the pipe's midpoint for (a) small initial condition ($q_1 = 1e-5$); and (b) large initial condition ($q_1 = 5e-3$)

Next, the Hencky model is evaluated with flow velocity and rotational speed. As in section 4.3, the total deflection of the midpoint of the system is obtained for a given rotational speed. The two values of rotational speed (i.e. $\Omega = 5$ and $\Omega = 8$) previously used in section 4.3 for the Galerkin model are employed here for the Hencky model to facilitate a direct comparison. Fig. 14 presents the results of this comparison.

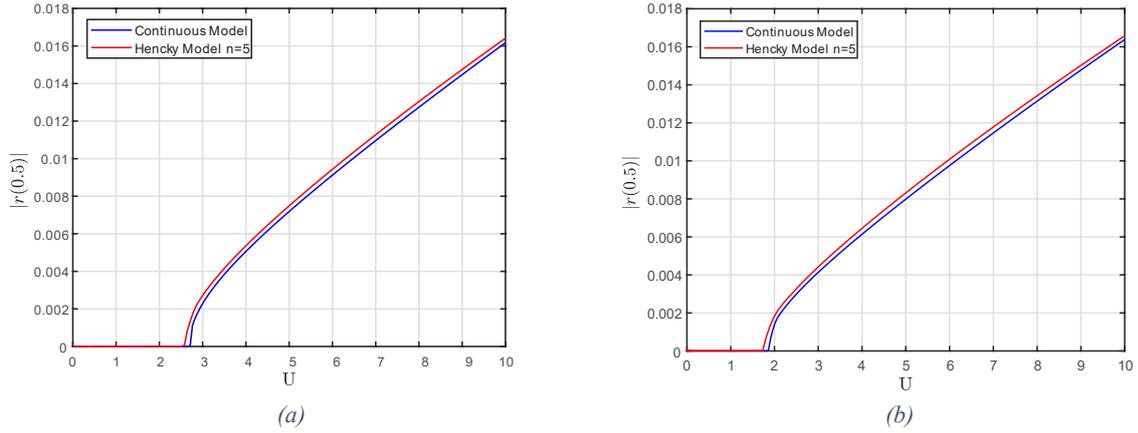

*Fig. 14. Total deflection of the midpoint of the system with respect to flow velocity; (a) $\Omega = 5$ and (b) $\Omega = 8$.*

The results confirm a strong agreement between the Hencky and Galerkin models, indicating that both yield reliable outcomes.

### 5.3 Determining the Optimal Number of Links

The reason for choosing this specific number of links is now explored by studying the effect of the number of links on the accuracy of the Hencky model's results. To determine the optimal number of links, the impact of both rotational speed and flow velocity on the solution is analyzed separately.

The effect of rotational speed on the system is studied when the flow velocity is set to zero. As shown in Fig. 15, even with $n = 3$, the Hencky model shows very good agreement with the Galerkin model. The Hencky model diverges from the zero-equilibrium position at $\Omega = 9.86$, which is very close to $\Omega = \pi^2$, obtained using the Galerkin method in Section 4.2. Therefore, it is concluded that the number of links has no significant effect on the accuracy of the solution as the rotational speed increases.

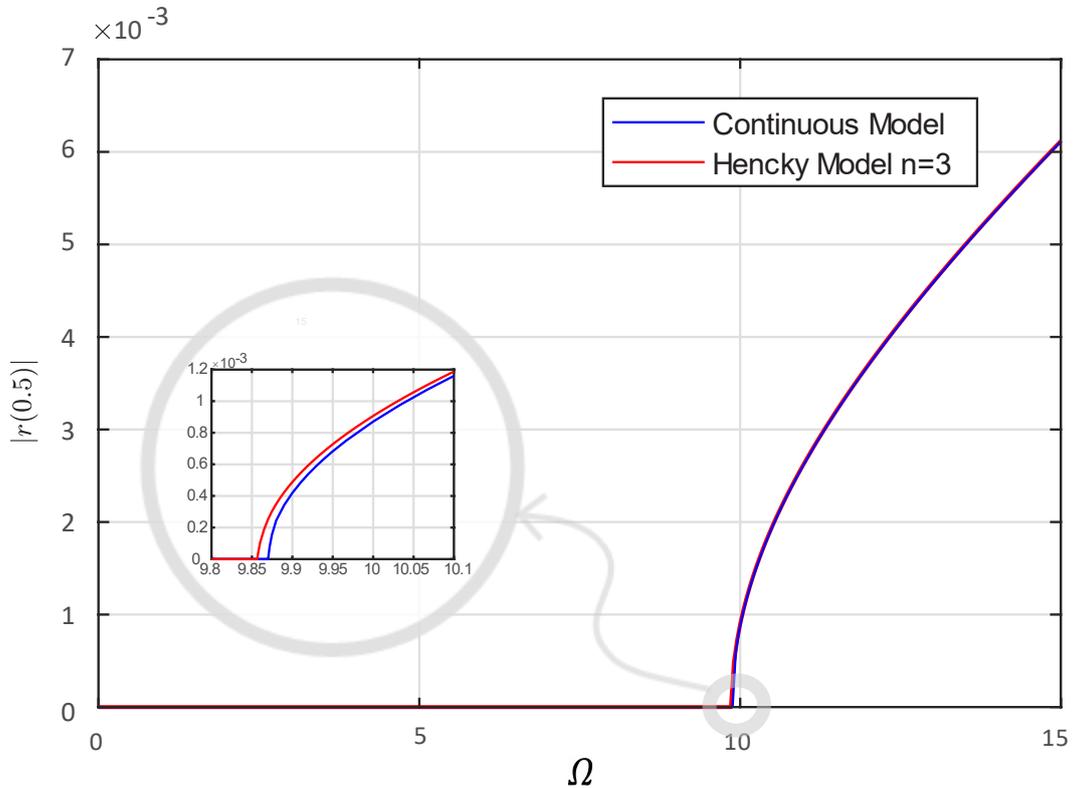

*Fig. 15. Bifurcation diagram of Poincare points of the pipe's midpoint*

When considering nonzero flow velocity, the impact of the number of links becomes more apparent. Fig 16 shows that, unlike rotational speed, increasing the flow velocity requires a greater number of links for the Hencky model to achieve accurate results. This is evident from the critical flow velocities: the 3-link Hencky model has a critical speed of about 3, the 5-link model about 3.08, and the 7-link model about 3.12. However, given that the computation cost of the 7-link model is much higher than the 5-link model, while their critical speeds are quite similar, it is concluded that $n = 3$ is the optimal number of links.

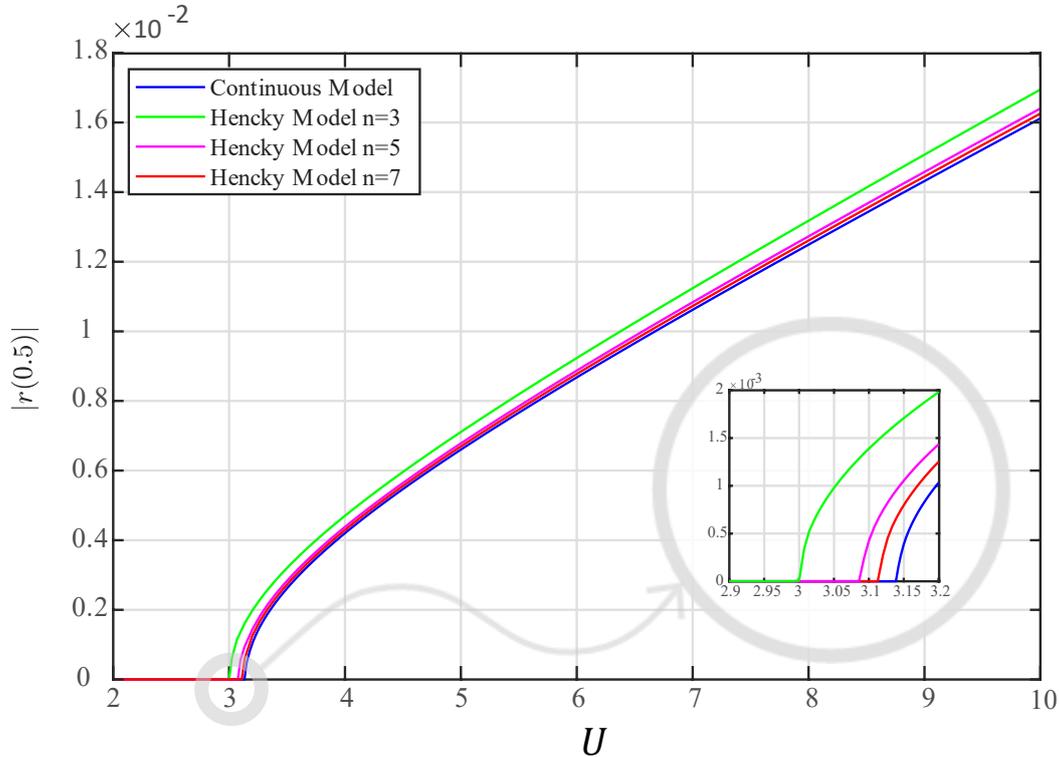

Fig 16. Bifurcation diagram of Poincare points of the pipe's midpoint

# 6. Conclusion

The nonlinear dynamics of a fluid-conveying pipe rotating with constant velocity about its longitudinal axis were analyzed in this study. By accounting for the stretching effect due to the pinned-pinned support, the nonlinear equation of motion was derived based on the Euler-Bernoulli beam theory. The internal damping, represented by the Voigt-Kelvin model, was included to capture the viscoelastic behavior of the pipe material. The equation of motion was discretized using the Galerkin method, and the system's dynamics was investigated through direct numerical simulations and frequency analysis. The following significant conclusions were drawn:

- Comparing the whirling frequencies in both the inertial and rotating frames revealed that describing the motion in the rotating frame alters the frequencies: as the rotational speed increases, the frequencies decrease.

- Rotational speed has a significant effect on the system's critical speed due to the presence of rotational damping. As the rotational speed increases, the system loses stability at progressively lower flow velocities.

- Post-instability solutions manifest as a non-zero equilibrium point representing the pipe's deflection, influenced by both flow velocity and rotational speed. Flow velocity has a greater influence compared to rotational speed. These non-zero equilibrium points are described in the rotating frame, whereas in the inertial frame, the corresponding result is a forward whirling motion with the angular frequency matching with the rotational speed.

- The stability region on the map of control parameters (flow velocity and rotational speed) is bounded by the critical flow velocity and critical rotational speed, forming a quarter-ellipse, with the semi-major axis ($\pi$) corresponding to the flow velocity and the semi-minor axis ($\pi^2$) corresponding to the rotational speed.

- A Hencky bar-chain model was developed to validate the results. The comparison between the Hencky and Galerkin models demonstrated strong agreement, confirming that both approaches yield reliable outcomes.

## Author contribution statement

Alis Fasihi: Conceptualization, Investigation, Methodology, Validation, Software, Formal Analysis, Visualization, Writing ‐ Original draft preparation, Writing ‐ Reviewing and Editing; Grzegorz Kudra: Conceptualization, Investigation, Methodology, Validation, Software, Formal Analysis, Writing ‐ Reviewing and Editing, Supervision; Maryam Ghandchi Tehrani: Supervision, Investigation, Formal Analysis; Jan Awrejcewicz: Supervision, Investigation, Formal Analysis.

## Acknowledgements

This work has been supported by the National Science Center, Poland under the grant PRELUDIUM 22 no. 2023/49/N/ST8/01957. This article was completed while the first author, Ali Fasihi, was a doctoral candidate at the Interdisciplinary Doctoral School at Lodz University of Technology, Poland.

For the purpose of Open Access, the authors have applied a CC-BY public copyright license to any Author Accepted Manuscript (AAM) version arising from this submission.

## Declaration of competing interest

The authors certify that no personal relationships or known competing financial interests could have appeared to influence the research presented in this paper.

## Data availability

Data will be provided upon request.

## Supplementary material

Supplementary data to this article can be found online.